\renewcommand\d{\delta}

	\newcommand\f{\phi}

	\renewcommand\o{\omega}
	
	\newcommand\D{\Delta}
	\newcommand\G{\Gamma}

	\newcommand{\diracslash}[1]{#1\llap{/\kern2pt}}
	
	\newcommand{\be}{\begin{equation}}
	\newcommand{\ee}{\end{equation}}
	\newcommand{\bea}{\begin{eqnarray}}
	\newcommand{\eea}{\end{eqnarray}}
	\newcommand{\ba}[1]{\begin{array}{#1}}
		\newcommand{\ea}{\end{array}}
	\newcommand{\bep}{\begin{pmatrix}}
		\newcommand{\eep}{\end{pmatrix}}
	
	\newcommand{\bt}{\begin{tabular}}
		\newcommand{\et}{\end{tabular}}

	\newcommand{\beas}{\begin{eqnarray*}}
		\newcommand{\eeas}{\end{eqnarray*}}

	    \documentclass[%
	    reprint,
	    amsmath,amssymb,
	    aps,
	]{revtex4-1}
	    \usepackage{amsmath}
	    \usepackage{cleveref}
	    \usepackage{amsfonts}
	    \usepackage{graphicx}
	    \usepackage{dcolumn}
	    \usepackage[utf8]{inputenc}
	    \usepackage{bm}
	\usepackage{multirow}
	\usepackage[T1]{fontenc}
	\usepackage{lineno}
	    \usepackage{titlesec}
	  \usepackage{easyReview}

	    \titlespacing{\subsubsection}{2pt}{\parskip}{-\parskip}
	    
\begin{document}
	
	    	\title{$\phi$ meson in nuclear matter and atomic nuclei}
	    	
	    	\author{Arpita Mondal}
	    	\email{arpita.mondal1996@gmail.com}
	
	   	\author{Amruta Mishra}%
	   	\email{amruta@physics.iitd.ac.in}

	    	\affiliation{%
	    		Department of Physics, Indian Institute of Technology Delhi, Hauz Khas, New Delhi 110016, India\\
	    	}
	
	
	    	\begin{abstract}
The properties (masses and decay widths) of the $\phi$ meson are investigated in nuclear matter from the $\phi$ meson self-energy, using the tree-level $\phi K\bar{K}$ Lagrangian, and, incorporating in-medium masses of (anti)kaons calculated within the quark meson coupling (QMC) model.
These mass shifts and decay widths are incorporated in the Breit-Wigner spectral function of the $\phi$ meson to calculate the production cross-section of $\phi$ in asymmetric nuclear matter.
Considerable modifications to the production cross-section are observed at normal nuclear matter density, driven by the in-medium mass reduction and the increase in the decay width of $\phi$ meson.
The potential experienced by $\phi$ meson in nuclear matter is used to study the possibility of formation of the $\phi$ mesic bound state with atomic nuclei.
We explore the potential formation of $\phi$-mesic bound states in ${\rm{^{4}He}}$, ${\rm{^{12}C}}$, ${\rm{^{16}O}}$, ${\rm{^{40}Ca}}$, ${\rm{^{90}Zr}}$, ${\rm{^{197}Au}}$ and ${\rm{^{208}Pb}}$ nuclei by investigating their binding energies and absorption widths based on the corresponding $\phi$-nucleus potentials.
Our study shows shallow bound states with the light nuclei and deeply bound states in heavy nuclei. 
Among the investigated nuclei, a particularly distinct signal for a $\phi$-mesic bound state is identified in ${\rm{^{16}O}}$, suggesting its potential experimental observability.
The work provides valuable insights into $\phi$ meson interactions in infinite nuclear matter and the potential formation of exotic $\phi$-mesic nuclear states, offering promising probes for strongly interacting matter in the upcoming experiments at J-PARC, JLab, and ${\rm{\bar{P}}ANDA}$@FAIR physics program.
	    	\end{abstract}
	    	\maketitle
	    	\section{\label{introduction}Introduction}
	    	The properties of light vector mesons has long been a focus of experimental and theoretical research, where the properties of these mesons have significant relevance in the branch of QCD (Quantum Chromodynamics) phenomenology.
	    	For instance, different extreme conditions of nuclear matter can significantly alter these properties, shedding light on the partial restoration of chiral symmetry in QCD.
	    	Because of the narrow natural width (4.26 MeV) \cite{pdg}, among all light vector mesons, $\phi$ meson has drawn particular interest.
	    	Unlike $\rho$ and $\o$ mesons, $\phi$ offers isolated resonance peak without much disturbance by complicated hadronic interactions, suggesting a distinct observation of the possible in-medium modification.
	    	
	    	Several experiments have indicated significant in-medium modifications of the $\phi$ meson properties.
	    	A mass reduction of around $3.4\%$ and an increase of the decay width of the $\phi$ meson by a factor of $3.6$ at normal nuclear matter density, $\rho_0$ as compared to the vacuum value \cite{pdg} were reported by KEK-PS E325 collaboration \cite{kek}.
	    	The recent study by ALICE collaboration \cite{alice} at LHC also supports the KEK-PS E325 result \cite{e16exp}.
	    	However, both the Laser Electron Photon (LEP) facility at SPring-8 \cite{leps} and Continuous electron beam accelerator facility (CEBAF) large acceptance spectrometer, CLAS, at JLab \cite{clas} reported a large in-medium $\phi N$ cross-section and estimated a broadening of $\f$ meson decay width of 23–100 MeV without any mass shift.
	    	A comparison with the model calculations, the ANKE-COSY Collaboration suggests the $\phi$ width to be as large as $\sim$ 50 MeV \cite{anke}.
	    	On the other hand, the model fit \cite{hades_model} of the HADES (High Acceptance DiElectron Spectrometer at the SIS18 at GSI) data \cite{hades} suggests an attractive $\phi N$ potential $\sim$ -(50–100) MeV at normal nuclear matter density.
	    	Hence, a unified consensus among the various experiments has yet to be achieved.
	    	Further comprehensions are needed to conclude the final status.
	    	Currently, J-PARC has put forward two experimental projects (i) using a 30 GeV proton beam (E16 Collaboration \cite{e16} and E88 Collaboration \cite{e88}) and (i) using a 1.1 GeV antiproton beam (E29 Collaboration) \cite{e29} to study the medium effects on the $\phi$ meson.
	    	
	    	The reported experimental evidence ensures that the $\phi$ does interact strongly with the nuclear environment, instead of being a nearly pure $s\bar{s}$ state.
	    	Since $\phi$ meson's vacuum spectral properties are characterized by its decay to $K\bar{K}$ ($\phi$ was discovered as a resonance in $K\bar{K}$ scattering experiment), the observed modifications are often attributed to changes in the $K\bar{K}$-loop contributions in dense nuclear matter.
	    	Experimentally, such in-medium properties are mostly studied through the spectral properties and cross-sections; and theoretically using various approaches.
	    	In the present work, we are interested in studying the influence of medium effects on the mass, partial decay widths, and the production cross sections (due to the scattering of $K^0 \bar{K}^0$ as well as $K^+K^-$) of the $\phi$ meson, using a phenomenological $\phi K\bar{K}$ Lagrangian incorporating the in-medium $K$ and $\bar{K}$ meson masses calculated within a quark meson coupling (QMC) model \cite{me23}.
	    	The the inclusion of scalar iso-vector meson, $\d$, breaks the isospin symmetry for the masses of light quark and antiquark doublets, causing mass splitting between ($u,\;d$) as well as ($\bar{d},\;\bar{u}$).
	    	As a consequence, the considered mesons exhibit mass splittings within the isodoublets of $K$( $K^+$, $K^0$) and $\bar{K}$($\bar{K}^0$, $K^-$) meson when embedded in asymmetric nuclear matter.
	    	
	    	As mentioned earlier, most of the experiments report a large in-medium broadening of the $\phi$ width or a small lifetime within such environment.
	    	This indicates a large absorption of $\phi$ within the nuclear radius, suggesting a strong attractive $\phi$-nucleus interaction. 
	    	For instance, analyses \cite{alice_hal} combining ALICE and HAL QCD Collaboration results provide evidence for the existence of a $\phi$-nucleus bound state.
	    	A recent analysis \cite{paryev} of HADES data further led to the real and imaginary parts of the $\phi$-nucleus potential of $\sim$ 50-100 MeV (for the $\phi$ momenta of $\sim$ 0.5–1.3 GeV) and $\sim$ 20–25 MeV ($\phi$ momentum of $\sim$ 0.8 GeV), respectively, indicates the feasible condition for the experimental observation of $\phi$-nucleus bound states even if the $\phi$ meson is not produced in recoilless kinematics.
	    	Moreover, in a reaction $\pi^- A \to \phi n X$, following the J-PARC E26 experiment, the slow $\phi$ mesons can be selected and the forward neutron measurement may lead to the potential observation of the $\phi$–nucleus bound states and $\phi$'s in-medium modification \cite{e26}.
	    	Notably, the J-PARC E29 collaboration is intended to study the in-medium mass modification of $\phi$ via the possible formation of the $\phi$-nucleus bound states, using the primary reaction $p\bar{p} \to \f\f$ \cite{e29}.
	    	The J-PARC E88 collaboration proposed \cite{e88} to study the in-medium modifications of $\phi$ inside the nucleus through $\phi \to K^+ K^-$ measurement with the E16 setup.
	    	Finally, JLab's proposal to study $\phi-{\rm{^{4}He}}$ with 12 GeV upgrade offers additional avenues for such investigation \cite{jlab}.
	    	In this paper, we report our studies on the possible formation of $\phi$ mesic nuclei bound states with ${\rm{^{4}He}}$, ${\rm{^{12}C}}$, ${\rm{^{16}O}}$, ${\rm{^{40}Ca}}$, ${\rm{^{90}Zr}}$, ${\rm{^{197}Au}}$, and ${\rm{^{208}Pb}}$ nuclei.
	    	Especially, we try to verify such possibilities and draw the experimental feasibility of the observation of those states.
	    	
	    	The present work has two aims: (i) $\phi$ in symmetric as well as asymmetric nuclear matter, and (ii) $\phi$ in symmetric as well as asymmetric nuclei.
	    	One may hope that the results obtained in the present work can be useful in planning these experiments.
	    	The paper is organized as follows.
	    	In Sec. \ref{matter} we present the theoretical base to study medium modifications of masses and decay widths of $\phi$ meson in nuclear matter and their effects on production cross-section.
	    	We also study the possibility of the $\phi$ meson bound within atomic nucleus.
	    	Then we present the results and discussions, starting with the discussions on the fitting of the parameters and the properties and production cross-sections of $\phi$ meson in symmetric and asymmetric nuclear matter in \ref{NMR} and \ref{PCR}, while Sec. \ref{FN} analyzes the formation of $\phi$-nuclei bound states using mesic-nuclei potentials.
	    	Finally, Sec. \ref{summary}, concludes with a summary of our findings in the current study.
	\section{$\phi$ meson in nuclear matter and atomic nuclei}
		\label{matter}
In this section, we discuss the properties of the $\phi$ meson within symmetric (SNM) as well as asymmetric (ANM) nuclear matter, especially how the mass and the decay widths receive medium modifications and their subsequent impact on production cross sections of $\phi$ mesons.
However, within the QMC model, $\phi$ does not interact directly with nuclear matter due to the absence of light (anti)quark constituents.
Therefore, one possibility of interaction is through the excitation of the intermediate state hadrons, which contain light (anti)quarks.
Considering the branching ratios of the $\phi$ meson, the intermediate kaon loop emerges as the most likely mechanism governing the $\phi$-nucleon interaction within nuclear matter. 
The $K\bar{K}$ decay channel accounts for approximately 83$\%$ of the total decay width of the $\phi$ meson \cite{pdg}.
Thus, the $\phi$ meson receives medium modifications through the kaon channels during such interaction.
The medium dependences of the open strange mesons $K$ and $\bar{K}$ have been computed within a quark meson coupling (QMC) model in Ref. \cite{me23}, where the incorporation of the scalar-isovector $\d$ meson breaks the isospin symmetry of the (anti)kaon isodoublets, leading to splitting within the masses of the isodoublets of $K$ and $\bar{K}$ mesons.
	\subsection{Masses and decay widths of $\phi$ mesons in nuclear matter} \label{MD}
\begin{figure}[th]
		\centerline{\includegraphics[width=9.5cm]{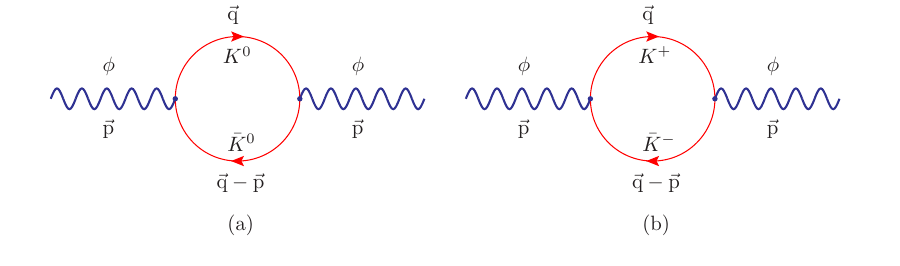}}
		\caption{\raggedright{Diagrams contributing to the $\phi$ self energy: (a) $K^0\bar{K}^0$ and (b) $K^+K$ loops. }}
		\label{loop}
	\end{figure}

	In the present study, we adopt a phenomenological Lagrangian describing the $\phi K \bar{K}$ interaction and investigate the in-medium mass and decay widths of $\phi$ meson.
	We calculate the decay widths due to the (i) $\phi \to K^0 \bar{K}^0$ and (ii) $\phi \to K^+ K^-$, accounted through kaon-antikaon loops, shown in Figs.\ref{loop}(a-b), with the following tree-level interactions,
	\onecolumngrid
\bea
\label{lag2}
\mathcal{L}_{\phi \to K^0 \bar{K}^0} &=& -i g_{\phi K^0 \bar{K}^0} \phi^\mu \Big(\bar{K}^0 \partial_\mu K^0 - K^0 \partial_\mu \bar{K}^0 \Big)~,\\
\label{lag1}
\mathcal{L}_{\phi \to K^+K^-} &=& -i g_{\phi K^+ K^-} \phi^\mu \Big( K^- \partial_\mu K^+ - K^+ \partial_\mu K^- \Big) ~,
\eea
where the coupling constants $g_{\phi K^0 \bar{K}^0}$ and $g_{\phi K^+ K^-}$ are fitted from the decay widths of $\phi$ to $K^0\bar{K}^0$ and $K^+K^-$, in vacuum \cite{pdg}.
With the four-momentum $p = (m_\phi,\vec{0})$ of the decaying particle, the $\phi$ meson self-energies associated with the loops, have the generic form
\bea
\label{selfE2}
\Pi_i^*(p)  
&=&i \frac{8}{3} g_i^2 \int \frac{d^4q}{(2\pi)^4}{\vec{q}}~^2 \frac{1}{(q^2-m_{K_i}^{*^2}+i\epsilon)}\frac{1}{((q-p)^2-m_{\bar{K_i}}^{*^2}+i\epsilon)}~,
\eea
where the superscript `$*$' denotes the in-medium quantities and $i$=1 and 2 refer to $K^0\bar{K}^0$ and $K^+K^-$ loops, and, $m_{K(\bar{K})_i}^*$ refers to $m_{K^0(\bar{K}^0)}^*$ and $m_{K^+(K^-)}^*$ respectively.
The imaginary part of $\Pi_i^*(p)$ contributes to the respective decay widths of the $\phi$ meson and the real part contributes to the medium modification of the mass of the $\phi$ meson through the in-medium $\phi$ propagator at rest,
\bea
\label{prop}
D_i (\o_i, |\vec{p}|\to 0,\rho) = \frac{1}{{\o_i}^2 - {m_{\phi_i}^0}^2 - \Pi_i(\o_i, |\vec{p}|\to 0,\rho)}
\eea
with the effective energy $\o_i = \sqrt{|\vec{p}|^2+{m^*_{\phi_i}}^2}$ and the in-medium $\phi$ mass, $m^*_{\phi_i}$, which can be solved by the dispersion relation,
\bea
\label{mass2}
m_{\phi_i}^{*^2}& - &\left(m_{\phi_i}^{0}\right)^{2} - {\rm{Re}}\Pi_{i}^*(m_{\phi_i}^{*^2}) = 0,
\eea
for the loop $i$ ($i$=1 and 2 for $K^0\bar{K}^0$ and $K^+K^-$), where, the bare mass, $m_{\phi_i}^0$, is fixed using Eq. (\ref{mass2}) in vacuum.
The real part  \cite{kreinplb11} of the self-energy takes the following form,
\bea
\label{re_selfE2}
{\rm{Re}}\Pi^*_{i}&=&-\frac{4}{3}g_{i}^{2}\, \mathcal{P}\!\!
\int^{\Lambda_K}_{0} \frac {d^3q} {(2\pi)^3} {q}^{2}\left( \frac{\Lambda^2_K+m_{\phi_i}^{*^2}}{\Lambda^2_K+4E_{K_i}^{*^2}}\right)^2 \left( \frac{\Lambda^2_K+m_{\phi_i}^{*^2}}{\Lambda^2_K+4E_{\bar{K}_i}^{*^2}}\right)^2\frac{(E^*_{K_i}+E^*_{\bar{K}_i})}{E^*_{K_i} E^*_{\bar{K}_i} ((E^*_{K_i}+E^*_{\bar{K}_i})^2-m_{\phi_i}^{*^2})}~,
\eea
where, $\mathcal{P}$ denotes the Cauchy principal value of the self-energy integral, $E^*_{K(\bar{K})_i}=({q}^{\,2}+m_{K(\bar{K})_i}^{*^2})^{1/2}$, and $\Lambda_{K}$ is the cutoff momentum parameter used to regularize the UV-divergence \cite{Ashok} appearing in the self-energy integrals.
The decay width is related to the imaginary part of the self-energy and is given as,
\bea
\label{Im_selfE2}
\Gamma_{\phi_i}^* = -\frac{1}{m_{\phi_i}^*} {\rm{Im}} \Pi^*_i = -\frac{2}{3\pi
} \frac{g_i^{2}|\vec{q_i}|^3}{m_{\phi_i}^{*}} ,
\eea
where,
\bea
|\vec{q_i}| = \frac{1}{2m_{\phi_i}^*} \left[(m_{\phi_i}^{*^2}-(m_{K_i}^{*}+m_{\bar{K}_i}^{*})^2)(m_{\phi_i}^{*^2}-(m_{K_i}^{*}-m_{\bar{K}_i}^{*})^2)\right]^{1/2}
\eea
is the magnitude of 3-momentum of the outgoing $K(\bar{K})$ meson.
	\vspace{0.6cm}
    \twocolumngrid
	\subsection{Production cross-sections of $\phi$ meson}\label{PC}
\begin{figure}[th]
		\centerline{\includegraphics[width=9.5cm]{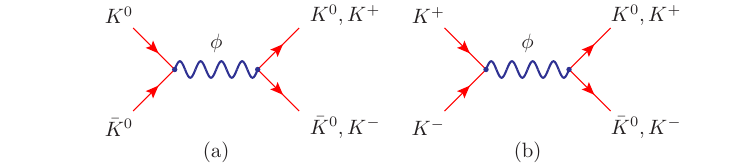}}
		\caption{\raggedright{Feynman diagrams associated with (a) $K^0\bar{K}^0 \allowbreak\to K\bar{K}$ and (b) $K^+K^- \to K\bar{K}$ processes.}}
		\label{scatt}
	\end{figure}

	The medium dependencies in turn influence the cross-section of the respective process, i.e., the $\phi$ meson is created from the scattering of (or decays to) particles in different channels ($K^0\bar{K}^0$ and $K^+K^-$).
	The probable processes are depicted in Fig.\ref{scatt}(a-b).
	The production cross-section of the $\phi$ meson \cite{Mishra21,Mishra24}, for the $i^{\rm{th}}$ initial-state, of $K^0\bar{K}^0$ and $K^+K^-$, represented by $i$=1 and 2, respectively, is 
	\bea
	\label{cros}
	\sigma_i ( K \bar{K} \to \f; M^2)=6 \pi^2 \frac{{\Gamma_{\phi_ i}^*}}{|\vec{q}_i|^2} A_{\phi_i}(M^2),
	\eea
	with M is the invariant mass and $A_{\phi_i}(M^2)$ is the relativistic Breit Wigner spectral function of
	the $\phi$ meson, which consists of real and imaginary parts of the self-energies through the following expression,
	\bea
	\label{spec}
	A_{\phi_i} (M^2)=C_i \frac{2}{\pi} 
	\frac {M^2 \Gamma_{\phi}^{* tot}}{(M^2-{m_{\phi_i}^*}^2)^2(M\Gamma_{\phi}^{*tot})^2},
	\eea
	where the in-medium total decay width of $\phi$ meson is given by the sum of the obtained partial decay widths, $\Gamma_\phi^{*tot}$ = $\sum_{i=1,2} \Gamma_{\phi_i}^*$ and $C_i$ is the constant, which is determined for the particular process by the normalization condition,
	\bea
	\label{norm}
	\int _0 ^{\infty} A_{\phi_i} (M^2) d M=1.
	\eea
	Notably, the $\phi$ meson gets corrections through the diagrams shown in Fig.\ref{loop}(a,b).
	The spectral function contains the medium properties i.e., density and isospin asymmetry for the present work.
	Thus, it carries all the information about the propagation of the mesonic excitation within the medium.
	\subsection{$\phi$ meson in atomic nuclei}
	\label{finite}
	As indicated in several studies, if mesons are produced in low momentum, an appreciable fraction of the produced mesons are expected to decay inside the nucleus, causing the formation of meson-nuclear bound states.
	This section lays out the potentials that a $\phi$ meson is experiencing while trapped inside a nucleus and the probable formation of the bound states with different nuclei.
	Since the $\phi (s\bar{s})$ meson consists of the same flavor of quark and antiquark, they feel equal and opposite Lorentz vector potentials.
	Therefore, the only existing potential experienced by the meson is the scalar potential. However, as we already have seen in Ref. \cite{me24}, all the potentials are correlated, but weakly. In the present scenario, the contributing potential is,
	\begin{equation}
	\label{potential}
	V_{\phi_i}(r)= U_{\phi_i}(r)-\frac{i}{2}W_{\phi_i}(r),
	\end{equation}
	with, $U_{\phi_i}(r)$ = $\Delta m_{\phi_i}(\rho(r))$$\equiv$ $m^{*}_{\phi_i}(\rho(r))-m_\phi$  and $W_{\phi_i}(r)$ = $\Gamma_{\phi_i}^*(\rho(r))$ are, respectively, the $\phi$-meson mass shift and decay width considering the $i^{th}$ loop contribution, where $i$=1(2) corresponds to $K^0\bar{K}^0 (K^+K^-)$ loop,
	Here, the nucleon density distribution $\rho(r)$ for the particular nucleus is obtained through the self-consistent calculation using the QMC model, as shown in Ref. \cite{me24} with $r$ is the distance from the center of the nucleus.
	With a careful inspection of the effective potential, the existence of the meson-nucleus bound states can be estimated naively.
	Using the $\phi$-meson-nucleus potentials we calculate the binding energies and absorption widths of $\phi$ meson in the nucleus by solving the following relativistic Klein-Gordon equation \cite{cobos17},
	\begin{equation}
	\label{kg}
	\left(-\nabla^{2} + \mu^{2} + 2\mu V_{\phi_i}({r})\right)\Phi_i({r})
	= \epsilon_i^{2}\Phi_i({r}),
	\end{equation}
	for various nuclei, where $\mu$ is the reduced mass of the $\phi$-meson-nucleus system.
	The bound state energies ($\rm{B}_i$) and absorption widths ($\Gamma_i$) are obtained from the complex energy eigenvalue $\epsilon_i$ by ${\rm{B}}_i={\rm{Re}}(\epsilon_i)-\mu$ and $\Gamma_i=-2{\rm{Im}}(\epsilon_i)$, respectively for the particular process.
	A negative binding energy and a well-behaved eigenstate within the nucleus for the given potentials are the signatures of such an exotic state.
	\section{Results and Discussions}
	\label{R}
	\subsection{$\phi$ meson in nuclear matter} \label{NMR}
We first state the parameters of the model we have chosen in the present work before discussing the results.
Since we are using the obtained in-medium masses of open strange mesons within SNM and ANM for the isospin asymmetry parameter $\eta$=0 and $\eta$=0.5, respectively, using the QMC model, as reported in our previous work, the parameters for the purpose can be found in the same Ref. \cite{me23}.
However, in the present study, the coupling constants of the particular channels i.e., $g_i$'s ($\equiv g_{K^0\bar{K}^0},~ g_{K^+K^-}$) need to be fixed.
The parameters are adjusted through their corresponding decay widths in free space, as mentioned in Table \ref{param}.
\begin{table}[th]
	\centering
	\begin{tabular}{{ccccccc}}
		\hline
		$i$  & $\phi \to K_L^0 K_S^0$ & $\phi \to K^+ K^-$\\
		\hline
		${\G_{\phi_i}}/{\G_{\phi}^{tot}}$ &  33.9 $\%$ & 49.1 $\%$ \\
		$\G_{\phi_i}$ (MeV) &  1.440 & 2.086 \\
		$g_i$ & 3.3212 & 3.2281\\
		\hline
	\end{tabular}
	\caption{ \raggedright{Branching ratios \cite{pdg}, empirical decay widths, and the extracted coupling strengths for each $\phi$ meson decay channel to two pseudoscalar mesons.}}
	\label{param}
\end{table}
	Another crucial step is to determine the cutoff momentum $\Lambda_K$ of the form factor and to fix the corresponding bare mass value. Since the integral is divergent, we regularize the associated loop integral with a phenomenological form factor with a cutoff mass parameter $\Lambda_K$, following the works in Refs. \cite{cobos17,cobos17plb,leinweberprd01, kreinplb11,klingl97,klingl98,yasuiprd09,zemiepja21}.
	We incorporate a dipole type form factor $\Big(\frac{\Lambda_K^2 + m_{\phi_i}^{*^2}}{\Lambda_K^2 + 4E_{K(\bar{K})_i}^2}\Big)^2$ at each $\phi K \bar{K}$ vertex and the cutoff $\Lambda_K$ is also introduced in the upper integration limit.
	Since $\Lambda_K$ has a direct correspondence with the overlap region between the respective mesons, it needs to be constrained by maintaining the physical properties of the mesons and the kinematics of the system.
	Refs. \cite{yasuiprd09,zemiepja21}, provides a vivid explanation of choosing the cutoff parameters.
	Depending on all the constraints, the cutoff parameter $\Lambda_K$ = 2-4 GeV is preferred for the current study.
	Now, the bare mass is fixed by the coupling constant and the vacuum mass at the particular cutoff value using Eq. (\ref{mass2}).
\begin{table}[th]
	\centering
	\begin{tabular}{cccc}
		\hline
		$\rm{\Lambda_K}$ & \multicolumn{2}{c}{$m_{\phi_i}^0~\rm{(MeV)}$} \\
		(GeV)& ($i=1$)  &  ($i=2$)  \\
		\hline
		2 & 1074.0 & 1073.2& \\
		3 & 1133.6 & 1130.7& \\
		4 & 1215.2 & 1209.4& \\
		\hline
	\end{tabular}
	\caption{\raggedright{Cutoff parameter $\Lambda_K$ dependence of the bare mass $m_{\phi_i}^0$. We show the variation of the bare mass calculated for the  neutral ($i=1$) and charged ($i=2$) kaon loop, which is minute but fluctuating.}}
	\label{bare}
\end{table}
	Table \ref{bare} shows the cutoff parameter dependencies of the bare mass $m_{\phi_i}^0$, indicating a small but increasing behavior of $m_{\phi_i}^0$ with $\Lambda_K$.
	As we can see, on each scale ($\Lambda_K$), considering the obtained bare masses, we notice that the physical $\phi$-mass is smaller than the bare masses, indicating an effective attractive in-medium scalar potential, which will be absorbed by the bare mass and move to its physical mass in vacuum.
	Notably, compared to the neutral kaon loop contributions, the obtained bare masses are on the lower side while considering the charged kaon loop, indicating that the attractive potential would value more in the case of `$i=1$' loop than of `$i=2$' loop.
	
\begin{figure}[th]
		\centerline{\includegraphics[width=4cm]{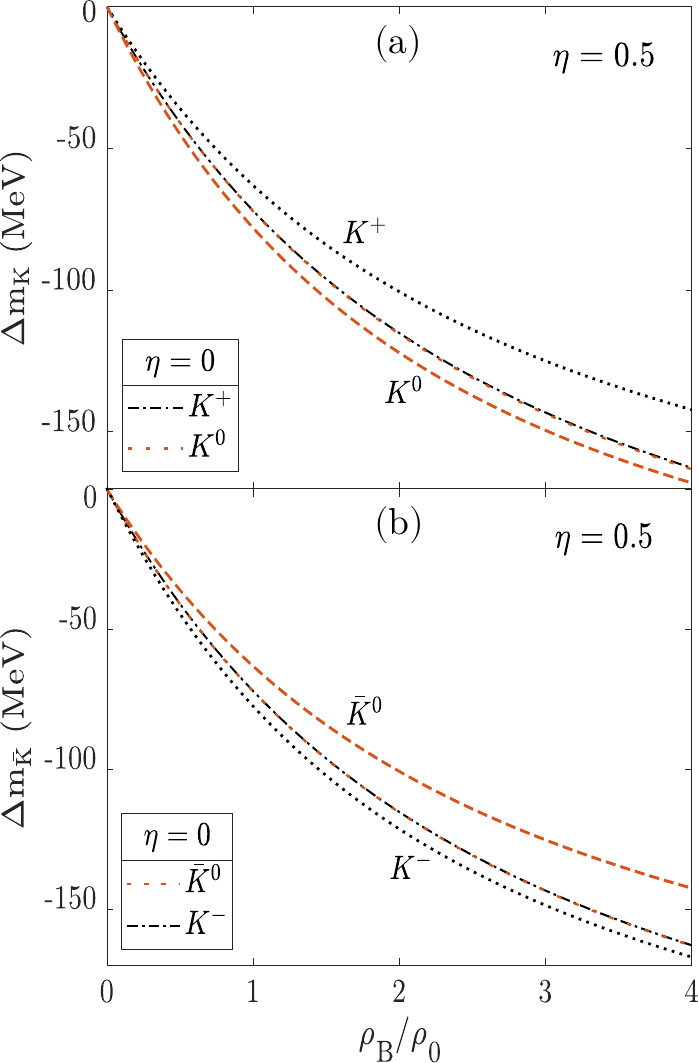}}
		\caption{\raggedright{The mass shifts of (a) $K$ (b) $\bar{K}$ mesons as a function of baryon density $\rho_B$ (in the units of $\rho_0$) \cite{me23}.}}
		\label{K}
	\end{figure}

	As discussed earlier, the $\phi$ meson properties receive medium modifications through the in-medium $\phi$ self-energy, arising from the interaction of $K$ and $\bar{K}$ mesons with the nucleons in the nuclear medium.
The properties of these mesons in nuclear matter undergo medium modifications due to the interactions of their constituent light (anti)quarks with the mean meson fields within the QMC model.
The inclusion of the $\d$ (scalar isovector) meson in the QMC model induces isospin symmetry breaking of the masses of light quark ($u,\;d$) and antiquark ($\bar{d},\;\bar{u}$) doublets, causing mass splittings within the isodoublets of $K(K^+, K^0)$ and $\bar{K}(\bar{K}^0, K^-)$ mesons in ANM, which are observed to increase with increasing baryon density.
A detailed analysis is presented in our previous work in Ref. \cite{me23}.
The behaviors of the mass modification ($\D_{m_{K(\bar{K})}}$ = $m_{K(\bar{K})}^* - m_{K(\bar{K})}$) of kaon and antikaon in symmetric as well as asymmetric nuclear matter with the asymmetry parameter $\eta=0$ and $\eta=0.5$, respectively, are illustrated in Fig.\ref{K}(a,b).

\begin{figure}[th]
		\centerline{\includegraphics[width=9cm]{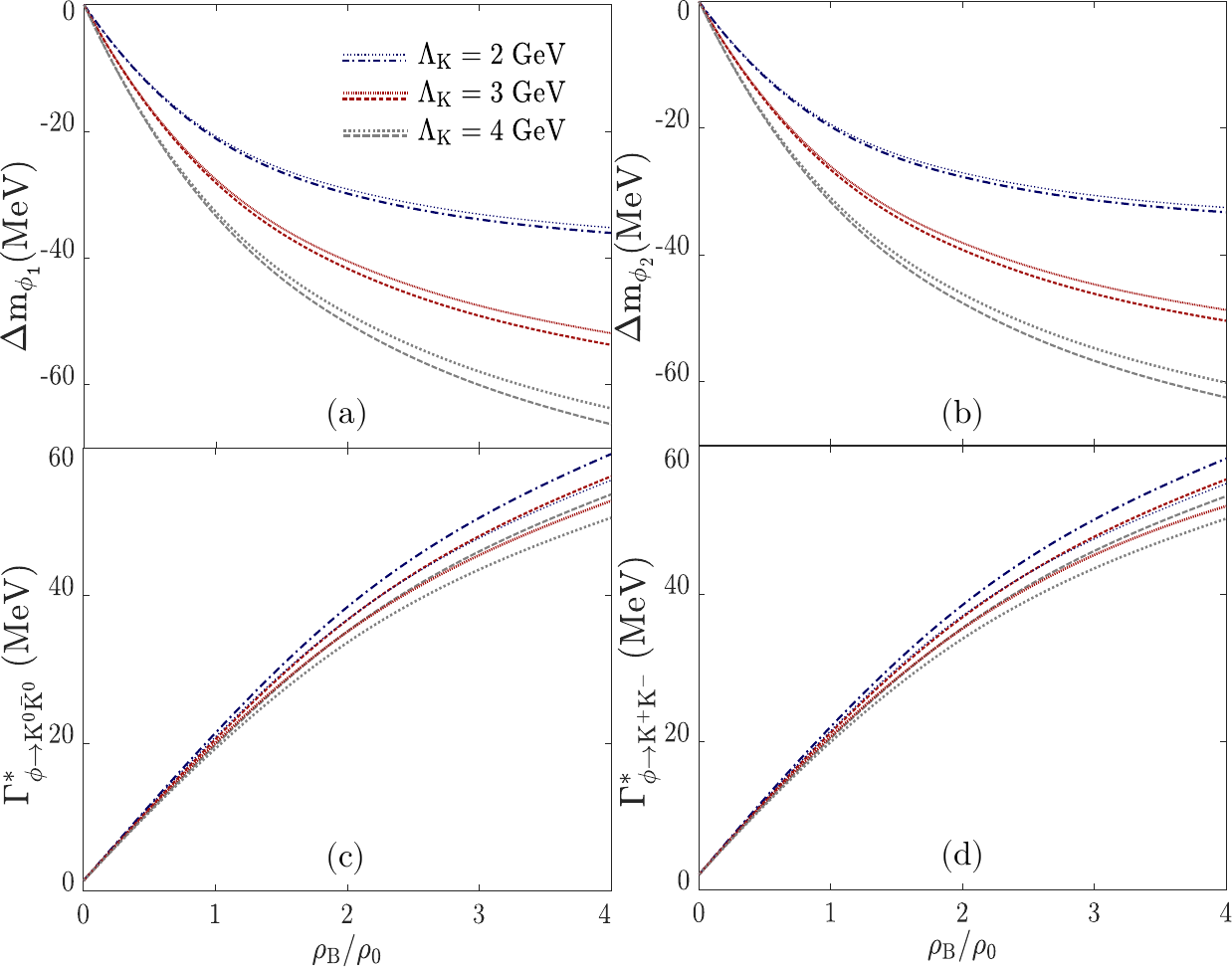}}
		\caption{\raggedright{(a-b) Mass shifts, $\D m_{\phi_i}$, and (c-d) decay widths of $\phi$ meson, $\Gamma_{\phi_i}^*$, considering the neutral (left panel; $i$=1) and charged (right panel; $i$=2) $K\bar{K}$ loop in asymmetric nuclear matter. The results are presented for different cutoff values and all the upper(lower) lines in the legend inset represent the results for $\eta$ =0.5($\eta$ =0).}}
		\label{md}
	\end{figure}

	We now analyze the mass shift, $\D_{m_{\phi_i}}$ (= $m_{\phi_i}^* - m_{\phi}$),  and the partial decay widths of the $\phi$ meson to open strange ($K^0 \bar{K}^0$ and $K^+K^-$) mesons by solving the regularized neutral as well as charged kaon loop integrals self-consistently, considering the respective in-medium kaon and antikaon masses.
	Figure \ref{md}(a-d) presents the in-medium $\phi$-mass and its partial decay widths in SNM as well as in ANM, for the densities up to $4\rho_0$ and a range of cutoff parameter $\Lambda_K$=2-4 GeV.
	As explored in Ref. \cite{me23}, the strong attractive scalar interaction leads to a reduction in kaon and antikaon masses, accompanied by comparatively smaller but significant splitting effects arising from the scalar-isovector interaction in ANM.
	Such modification in the (anti)kaon masses enhances the respective $K\bar{K}$-loop contribution in the nuclear medium relative to the vacuum, resulting a mass drop and the partial decay width broadening at finite densities, which increase as the density rises.
	For instance, considering `$i=1$' loop, $\Lambda_K$=2 GeV, and $\eta$=0, at $\rho_B = \rho_0$, the mass drop is $\sim$ 2.1\% with respect to its vacuum value of 1019.461 $\rm{MeV}$ whereas, it becomes around 3.5\% at $4\rho_0$;  the corresponding partial decay width has increased almost 13.7 (37.3) times from its vacuum value 1.44 MeV at $\rho_0$ ($4\rho_0$).
	However, as $\Lambda_K$ increases, the partial decay widths show a slow decrease with a steadily increasing mass drop.
	Under the same condition, at $\Lambda_K$= 3 GeV, the mass drop changes to $\sim$ 2.7\% (5.3\%) with almost 1.2 (3.5) MeV decrease in $\Gamma_{\phi_{1}}^*$ (from the case of $\Lambda_K$ = 2 GeV) at $\rho_0$($4\rho_0$).
	Due to the nonzero contribution of scalar-isovector interaction, the $\phi$ meson undergoes slightly smaller mass drop (and decay width) in ANM compared to SNM.
	Within the same scenario, at $\Lambda_K$= 2 GeV, as we move to ANM with the nonzero isospin asymmetry parameter, $\eta$=0.5, from SNM, the in-medium $\phi$-mass drop decreases by 0.4(0.9) MeV and the partial decay width decreases by 0.5(3.2) MeV at $\rho_0$($4\rho_0$).
	Therefore, these differences increase with the increasing density.
	However, even if the in-medium mass is increasing with the rising $\Lambda_K$, the values of the partial decay widths become smaller.
	The dependences of the obtained results on different parameters are summarized in Table \ref{sat_m}, where we observe that the neutral loop induces an overall smaller in-medium $\phi$ mass as well as a smaller decay width in comparison with the charged loop contribution.
	\begin{table}[th]
		\centering
		\begin{tabular}{ccccc}
			\multicolumn{5}{c}{$\rho_B = \rho_0$} \\ 
			\hline
			$\rm{\Lambda_K}$ & & 2 GeV & 3 GeV & 4 GeV \\
			\hline
			\multirow{2}{*}{$m_{\phi_i}^*~\rm{(MeV)}$}&($i=1$)& 998.3(998.7) & 991.4(991.9) & 986.1(986.8) \\
			&($i=2$)	& 999.7(1000.1) & 992.9(993.4) & 987.8(988.5) \\
			\hline
			\multirow{2}{*}{$\Gamma_{\phi_{i}}^*~\rm{(MeV)}$}& ($i=1$) & 19.8(19.3) & 18.5(18.1) & 17.6(17.2) \\
			&($i=2$)	& 20.5(20.0) & 19.3(18.9) & 18.5(18.0) \\
			\hline
			\multicolumn{5}{c}{$\rho_B = 2\rho_0$}\\
			\hline
			\multirow{2}{*}{$m_{\phi_i}^*~\rm{(MeV)}$}&($i=1$) & 989.7(990.4) & 977.8(979.0) & 969.2(970.7) \\
			&($i=2$)	& 991.8(992.4) & 980.2(981.3) & 971.9(973.3) \\
			\hline
			\multirow{2}{*}{$\Gamma_{\phi_{i}}^*~\rm{(MeV)}$}& ($i=1$)& 35.2(33.6) & 32.9(31.5) & 31.2(29.9) \\
			&($i=2$)	& 35.5(34.0) & 33.3(31.9) & 31.8(30.5) \\
			\hline
			\multicolumn{5}{c}{$\rho_B = 4\rho_0$}\\
			\hline
			\multirow{2}{*}{$m_{\phi_i}^*~\rm{(MeV)}$}&($i=1$) & 983.4(984.3) & 965.8(967.6) & 953.2(955.7) \\
			&($i=2$)	& 986.2(987.0) & 969.0(970.8) & 956.9(959.3) \\
			\hline
			\multirow{2}{*}{$\Gamma_{\phi_{i}}^*~\rm{(MeV)}$}&($i=1$) & 53.7(50.5) & 50.2(47.3) & 47.8(44.9) \\
			&($i=2$)	& 53.3(50.2) & 50.1(47.2) & 47.8(45.0) \\
			\hline
		\end{tabular}
		\caption{ \raggedright{Nuclear matter density and cutoff parameter $\Lambda_K$ dependence of the in-medium $\phi$ mass and its partial decay widths, considering $K^0\bar{K}^0$ ($i=1$) and $K^+K^-$ ($i=2$) loops are shown for SNM(ANM).}}
		\label{sat_m}
	\end{table}
As can be seen in Table \ref{sat_m}, compared to Ref. \cite{cobos17plb}, the obtained in-medium masses are lesser, depicting the influence of the scalar-isovector interaction.
However, compared to the work \cite{rajesh20}, where the (anti)kaon's masses are studied within the chiral effective model, the currently obtained masses are quite larger.
Similarly, the obtained partial decay widths are also larger compared to the widths obtained in Ref. \cite{Mishra21}, where the decay widths are computed using a field theoretic model of composite hadrons with quark (and antiquark) constituents with the in-medium (anti)kaon masses calculated within chiral SU(3) model.

\begin{figure}[th]
	\centerline{\includegraphics[width=4.5cm, height=6.8cm]{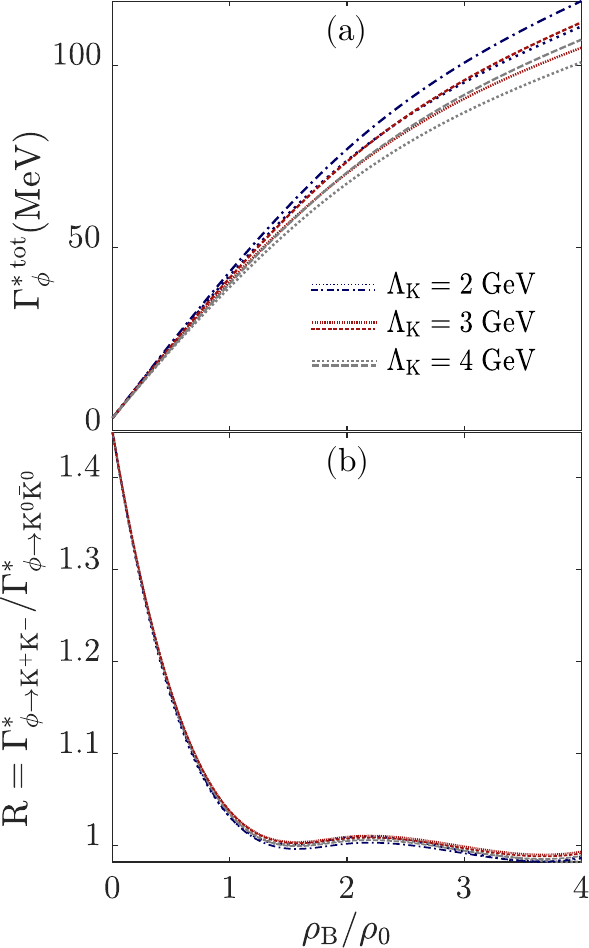}}
	\caption{\raggedright{(a) Total decay width of $\phi$ meson, $\Gamma_{\phi}^{*tot}$, and (b) variation of the ratio of partial decay widths of the two kaon channels, $R$, in SNM (lower lines in the legend inset) and ANM  (upper lines in the legend inset) are presented for different cutoff values.}}
	\label{d}
\end{figure}

The total decay width accounting both kaon channels, depicted in Fig.\ref{d}(a), demonstrates the in-medium broadening of the $\phi$ meson.
For instance, at $\rho_B = \rho_0$, the total decay width increases significantly, around 35–40 MeV for the whole range of $\Lambda_K$,  from its small vacuum width of 3.5 MeV \cite{pdg}.
The values are slightly higher, with a difference of $\sim$ 4 MeV, compared to the widths computed in QMC model \cite{cobos17plb}, without accounting the scalar-isovector interaction and these values increase as the density increases.
Independent of  $\Lambda_K$, at nonzero densities, the obtained decay widths are smaller in ANM than in SNM, and as the density increases, the difference between them increases (at $4\rho_0$ the difference becomes almost around 6 times than in $\rho_0$).
With the rise of $\Lambda_K$, the values start to decrease in both SNM and ANM, with a decreasing difference as well.
The obtained decay width at $\rho_0$ is larger as compared to the value of around 25 MeV in Ref. \cite{ko92} and, the value of around 22 MeV using a coupled channel approach \cite{oset01}; and even greater compared to the widths of $\sim$ 6 MeV in asymmetric nuclear matter obtained in Refs. \cite{Mishra21,rajesh20}, where in both works, the (anti)kaon mass modifications are estimated from the chiral SU(3) model.
However, the obtained width is very similar to the values of $\Gamma_\phi^{*tot} \simeq$ 45 MeV calculated in Ref. \cite{klingl98}.
As a consequence of such significant medium effects, the lifetime of the $\phi$ meson is reduced to around 4.4 $\rm{fm}$ at normal nuclear matter compared to its much longer vacuum lifetime of around 55 $\rm{fm}$, which is much bigger to observe any medium effects. 
This substantial reduction in lifetime, within the medium, indicates that $\phi$ mesons are likely to decay within nuclear dimensions \cite{klingl98} when produced in low-energy hadronic collisions.
Such scenario provides an opportunity to observe the medium modifications of the $\phi$ meson.
Furthermore, a small vacuum width of $\phi$ meson reflects that the kaon pairs have less available phase space, and thus $\phi \to K \bar{K}$ takes place barely above the threshold.
As the density increases, the decreasing in-medium mass (ensuring $M_\phi^* \geq 2m_K^*$) and the width broadening indicate the $\phi$ mesons are more likely to decay before freeze-out, and a lower yield of phi mesons can be detected through their decay into kaons.
Alternatively, such an enhancement in the decay width can also be attributed to the possibilities of inelastic $\phi N$ interactions, such as, $\phi N \to K \Lambda,~ K \Sigma$, etc. \cite{klingl98}.
	
   The subplot (b) of Fig.\ref{d} illustrates the ratio of the partial decay widths for the charged and neutral decay channels.
   The quantity represents the ratio of their respective branching ratios.
   In vacuum, this ratio is $\sim$ 1.45, indicating the breaking of the SU(2) isospin symmetry and a small electromagnetic contribution.
   Precisely, since the $d\bar{d}$ pair is heavier than the $u\bar{u}$ pair, the decay channel $\phi \to K^0 \bar{K}^0$ is relatively suppressed, thereby increasing the ratio above unity.
   At normal nuclear matter density ($\rho_0$), the ratio decreases to nearly 1.05 and approaches unity as the density is increased further. 
   This trend appears to be largely independent of the cutoff parameter.
   A simple explanation for this behavior might suggest that the $\phi \to K \bar{K}$ decay modes proceed through strong interaction dynamics at a good isospin limit.
   Moreover, a reduction in the vector boson mass within the medium is often associated with the restoration of chiral symmetry.
   However, Fig.\ref{K} indicates that the isospin symmetry breaking effect becomes pronounced at higher densities and similar behaviors are also illustrated in previous studies \cite{me23,santos,liu}.
   This suggests that the naive explanation might not fully capture the observed trend.
   In Ref. \cite{me23}, it was demonstrated that the splittings between the components of quark isodoublets and antiquark isodoublets are nearly identical, with these splittings increasing as the density rises.
   As shown in Fig.\ref{K}, the density-dependent effective masses of each kaon channel compensate one another in this ratio.
   For instance, at nonzero densities, the relations $M_{K^+}^*>M_{K^0}^*$ and $M_{\bar{K}^0}^*>M_{K^-}^*$ hold;
   and at high densities, the magnitude of the 3-momentum of $K^+(K^-)$ and $K^0(\bar{K}^0)$ are modified in such a way so that the ratio of $\Gamma_i^*$'s for the channels, $i$=1,2, approaches unity.
   This suggests that the ratio approaching unity at higher densities likely indicates that both decay channels gain roughly equal access to the phase space in this regime, leading to comparable branching fractions.
	
	In the next subsection, we discuss the production cross-section of the $\phi$ meson in SNM as well as in ANM, which are obtained from the in-medium masses of $\phi$ meson and the partial decay widths within the nuclear medium.
	\subsection{Production cross-section for $\phi$ meson} \label{PCR}
	\begin{figure*}[th]
		\centerline{\includegraphics[width=12cm]{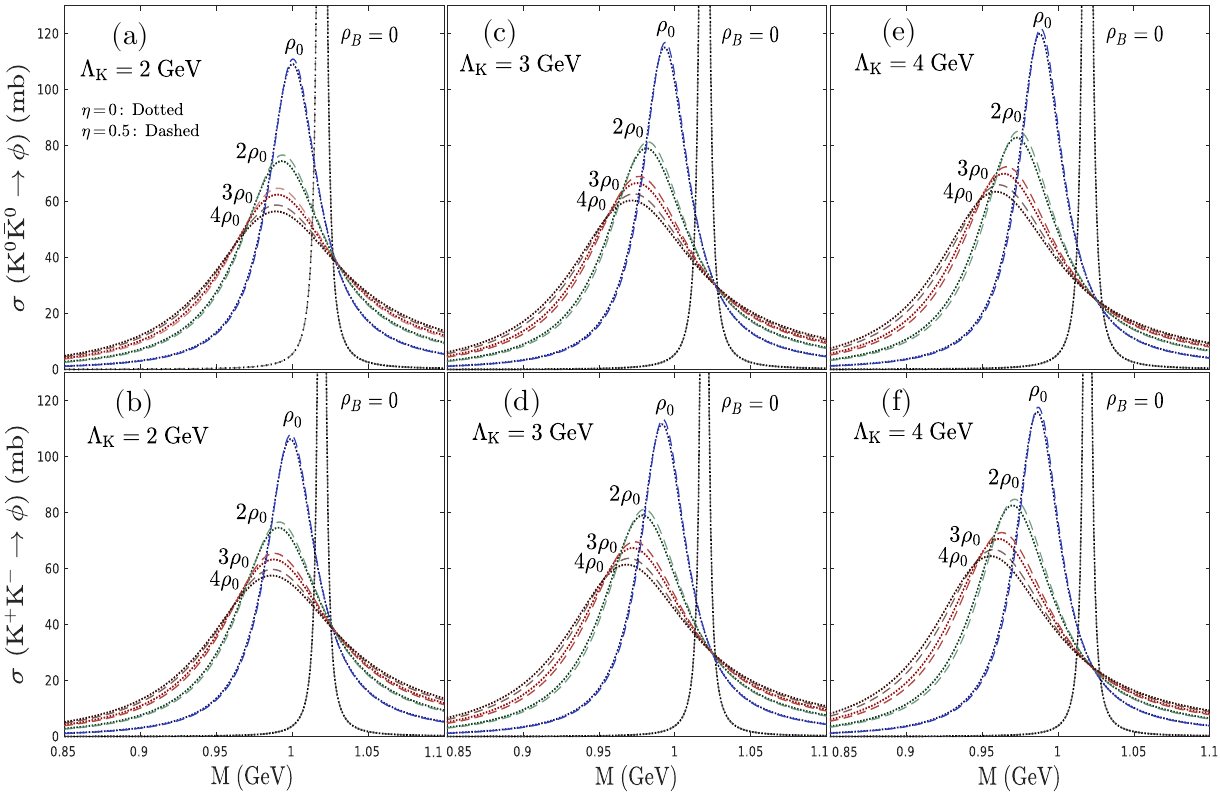}}
		\caption{\raggedright{Production cross-sections of $\phi$ meson due to scattering of $K^0\bar{K}^0$ (upper panel) and $K^+K^-$ (lower panel) mesons are plotted as functions of the invariant mass, for vacuum as well as for symmetric (dotted lines) and asymmetric (dashed lines) nuclear matter at $\rho_B = \rho_0, 2\rho_0, 3\rho_0, 4\rho_0$. We show the variations of the production cross-section with the different cutoff values (a-f).}}
		\label{cr}
	\end{figure*}
The production cross-section of the $\phi$ meson, arising from the $K^0 \bar{K}^0$ and $K^+ K^-$ two-body scattering processes, is calculated from its spectral function, which incorporates density and isospin asymmetry dependent effective masses and partial decay widths of $\phi$ meson, as obtained from the real and imaginary parts of the $\phi$ self-energy, respectively.
Figure \ref{cr} illustrates the production cross-sections of the $\phi$ meson in vacuum as well as at finite baryon densities $\rho_0$, $2\rho_0$, $3\rho_0$, and $4\rho_0$, for all the values of $\Lambda_K$, considered in this study.
These calculations are performed for SNM ($\eta = 0$) and ANM ($\eta = 0.5$).
	In vacuum, the cross-section peaks sharply around the pole mass of 1019.461 MeV. 
	As the energy approaches the resonance mass, the amplitude of the cross-section increases rapidly and then suddenly drops, producing a peak, centered around the pole.
	However, within the medium, the $\phi$ meson experiences an effective strong attractive interaction, resulting in a shift of the peak of the cross-section towards the lower invariant masses with reduced amplitudes.
	The peak is observed to broaden, and its height is observed to decrease with the increasing nuclear density.
	However, because of the relatively weak $\phi N$ interaction, only a small modification of the peaks is observed within the medium.
	In ANM, for $\rho_B$=$\rho_0$, the values vary between 108(111) and 118(122) mb, while for $\rho_B = 4\rho_0$, they range from 55(59) to 63(67) mb across the cutoff parameters $\Lambda_K$ = 2-4 GeV from the vacuum value of around 485(704) mb, considering `$i=1$'(`$i=2$') loop.
	Simultaneously, the pole position shifts to lower invariant masses with increasing density.
	As can be seen in Fig.\ref{cr}, the obtained peak heights are slightly smaller in SNM with comparatively lower invariant masses, which enhances at higher densities.
	This suggests that as baryon density increases and the isospin asymmetry decreases, the threshold energy for the creation of a $\phi$ meson decreases, enabling the creation of off-shell $\phi$ mesons at lower invariant masses.
	Moreover, a simultaneous decrease in kaon masses prevents the in-medium $\phi$ mass from exceeding the two kaon mass thresholds.
	The peaks corresponding to each density become more distinct with the higher values of cutoff parameter, and the isospin asymmetry effect is pronounced at higher values of density as well as $\Lambda_K$.
	This in-medium broadening of the resonance, results in low-mass dilepton enhancements below the $\phi$ mass serves as a key observable for detecting modifications in chiral properties.   
	Furthermore, in Fig.\ref{cr}, both channels exhibit similar behavior under the given nuclear matter conditions, with a slight difference between the peak positions where irrespective of $\Lambda_K$ values, the peak corresponds to the charged channel stays on a little right compared to the neutral channel peak.
	This shape change in the $\phi$ meson's spectral peak highlights the medium-induced modifications to its spectral properties; observing such shifts in the $\phi$ meson peak within dikaon spectra would require high-resolution beams. 
	\subsection{$\phi$ meson in atomic nuclei} \label{FN}
	In the present subsection, we study the possibility of $\phi$ meson bound in atomic nuclei.
	For this, the parameters considered are taken from Ref. \cite{me24}.
	As we discussed, the properties of $\phi$ meson within nuclear matter indicate the possibility of the formation of meson-nucleus bound states, where an appreciable fraction of produced mesons is expected to decay inside the nucleus.
	The nuclear environment facilitates the formation of such states.
	As studied earlier in ref. \cite{me24,tsushimaplb98}, under certain scenarios, the meson could be placed much closer to the nuclei, forming a deeply bound state, which allows to probe even subtle variations in medium properties.
	A precise measurement of their binding energies is crucial in understanding meson-nucleus interactions, the complex dynamics of strongly interacting systems, and the nature of these states.
	Therefore, in this section, we discuss the probable formation of $\phi$ meson-nucleus bound states with ${\rm{^{4}He}}$, ${\rm{^{12}C}}$, ${\rm{^{16}O}}$, ${\rm{^{40}Ca}}$, ${\rm{^{90}Zr}}$, ${\rm{^{197}Au}}$, and ${\rm{^{208}Pb}}$ nuclei, considering $\phi$ meson interaction with the nuclear environment through the neutral and charged kaon loop contributions.

\begin{figure}[th]
		\centerline{\includegraphics[width=8.2cm]{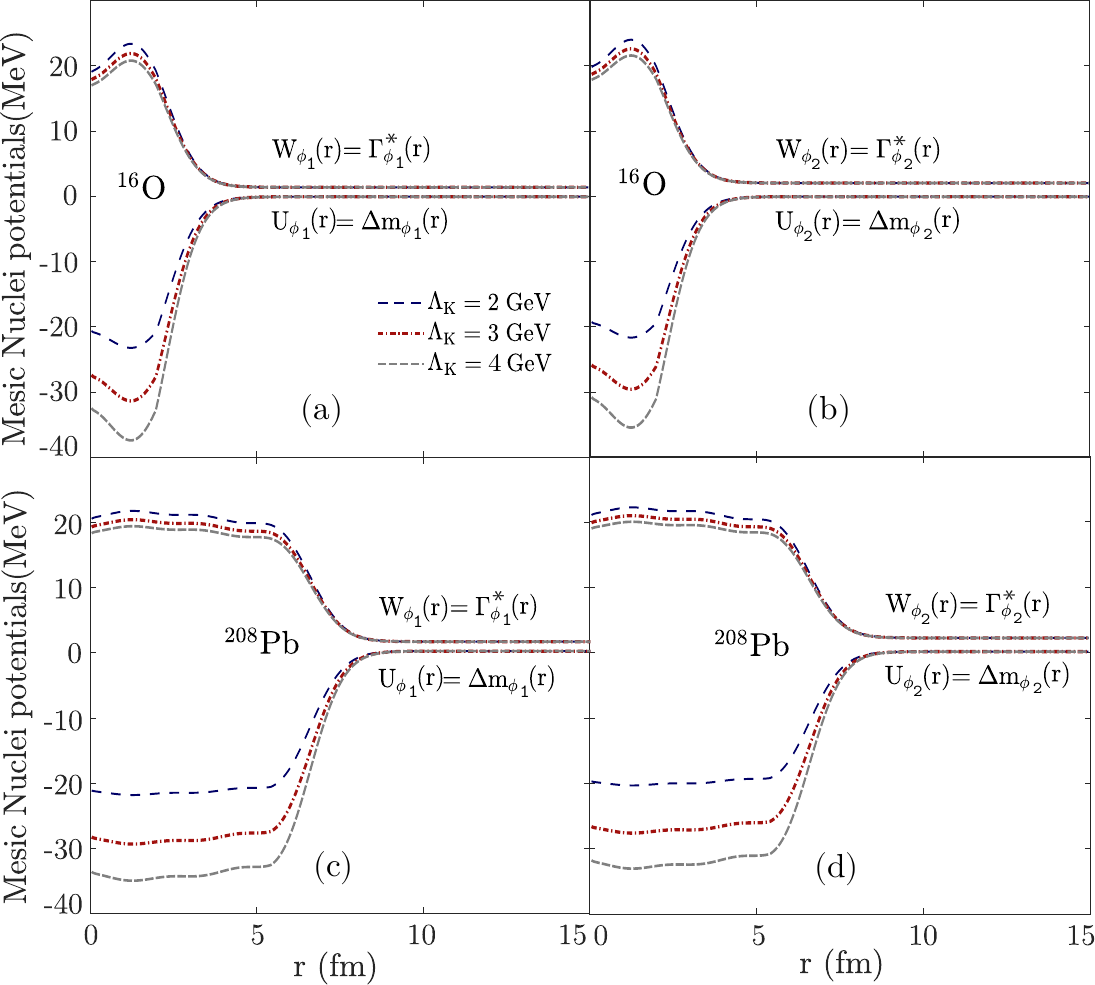}}
		\caption{\raggedright{Behaviors of real and imaginary parts of the mesic nuclei potential while a $\phi$ meson is captured inside an atomic nucleus, considering the neutral (left panel; $i=1$) and charged (right panel; $i=2$) kaon loop contributions. Results are presented for (a,b) symmetric (${\rm{^{16}O}}$) and (c,d) asymmetric (${\rm{^{208}Pb}}$) nucleus.}}
		\label{fn}
	\end{figure}

	We start with analyzing the mesic-nuclei potentials when the $\phi$ meson is captured inside a nucleus.
	As we have seen in ref. \cite{me24}, the existing potentials are largely drafted by the respective nuclear density distributions.
	Then, using the local density approximation we calculate the $\phi$-meson complex potentials for a particular nucleus.
	In Fig.\ref{fn}, we present the real and imaginary part of the $\phi$-meson complex potentials calculated for the seven nuclei for the considered range of the cutoff parameter $\Lambda_K$.
	Similar to the nuclear matter properties, the obtained depth of the real part of the potential, $U_{\phi_i}(r)$, is comparatively more sensitive to the cutoff parameter than the imaginary part $W_{\phi_i}(r)$.
	The obtained behaviors are illustrated in Fig.\ref{fn}, where we present the results for one isospin symmetric (${\rm{^{16}O}}$) and one isospin asymmetric (${\rm{^{208}Pb}}$) nucleus.
	From the behavior of the potentials, the plausible bound states can be speculated easily.
	For instance, as observed in Fig.\ref{fn}(c), for $\Lambda_K = 4~\rm{GeV}$ at normal nuclear matter density, the obtained potential of around $(-33 - i 9)$ MeV, indicating a bound state with approximately  $- 33$ MeV binding and a half-width of about 9 MeV.
	
\begin{figure}[th]
		\centerline{\includegraphics[width=5.5cm]{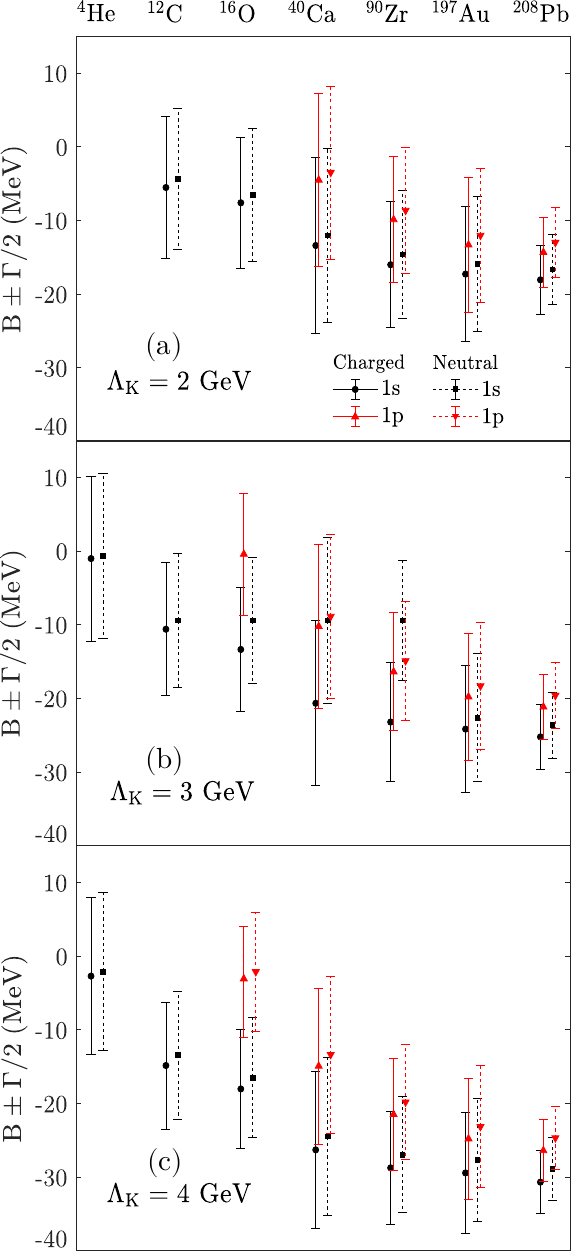}}
		\caption{\raggedright{Binding energies and absorption widths for different $\phi$ mesic-nucleus states. The variation of the bound states with the cutoff parameter is presented (a-f), considering both the `Neutral' and `Charged' kaon loop contributions. }}
		\label{be}
	\end{figure}

Using the obtained complex potentials, we estimate the binding energies and the absorption widths of probable $\phi$-mesic–nuclei states from the Klein-Gordon equation, while the $\phi$ meson is within the nuclei.
The calculated bound-state energies (${\rm{B}}_i$) and the half widths ($\Gamma_i/2$) for different nuclei, all the cutoff values, and both the neutral ($i=1$) and charged ($i=2$) kaon loop contributions are presented in Table \ref{B}.	
\begin{table}[th]
	\centering
	\begin{tabular}{{ccccccccccc}}
		\hline
		$\Lambda_K$ &&& 2 GeV & 3 GeV & 4 GeV &\\
		\hline
		${\rm{^{4}He}}$ & $1s$& B & $\times$($\times$) & -1.02(-0.64)& -2.70(-2.09)&\\
		&&$\Gamma/2$ & $\times$($\times$) & 11.21(11.21)& 10.68(10.72)&\\
		& $1p$& B& $\times$ ($\times$) & $\times$($\times$)& $\times$($\times$)&\\
		&&$\Gamma/2$ & $\times$($\times$) & $\times$($\times$)& $\times$($\times$)&\\
		\hline
		${\rm{^{12}C}}$ & $1s$& B & -5.53(-4.63) &-10.61(-9.41)&-14.84(-13.42)&\\
		&&$\Gamma/2$ & 9.59(9.63) & 9.02(9.09)& 10.68(10.72)&\\
		& $1p$& B& $\times$ ($\times$) & $\times$($\times$)& $\times$($\times$)&\\
		&&$\Gamma/2$ & $\times$($\times$) & $\times$($\times$)& $\times$($\times$)&\\
		\hline
		${\rm{^{16}O}}$ & $1s$& B & -7.60(-6.55) &-13.36(-12.02)&-18.02(-16.48)&\\
		&&$\Gamma/2$ & 8.94(8.99) & 8.41(8.50)& 8.03(8.15)&\\
		& $1p$& B& $\times$ ($\times$) & -0.42($\times$)& -3.06(-2.11)&\\
		&&$\Gamma/2$ & $\times$($\times$) & 8.30($\times$)& 7.90(8.03)&\\
		\hline
		${\rm{^{40}Ca}}$ & $1s$& B & -13.40(-12.04) &-20.66(-19.03)&-26.29(-24.45)&\\
		&&$\Gamma/2$ & 11.90(11.84) & 11.21(11.21)& 10.72(10.75)&\\
		& $1p$& B& -4.51(-3.49) & -10.22(-8.88)& -14.92(-13.36)&\\
		&&$\Gamma/2$ & 11.79(11.74) & 11.09(11.09)& 10.59(10.63)&\\
		\hline
		${\rm{^{90}Zr}}$ & $1s$& B & -16.01(-14.61) &-23.21(-21.59)&-28.73(-26.92)&\\
		&&$\Gamma/2$ & 8.57(8.64) & 8.07(8.18)& 7.71(7.84)&\\
		& $1p$& B& -9.88(-8.64) & -16.39(-14.90)& -21.48(-19.80)&\\
		&&$\Gamma/2$ & 8.51(8.59) & 8.02(8.12)& 7.65(7.78)&\\
		\hline
		${\rm{^{197}Au}}$ & $1s$& B & -17.28(-15.92) &-24.17(-22.61)&-29.42(-27.69)&\\
		&&$\Gamma/2$ & 9.15(9.20) & 8.62(8.71)& 8.23(8.35)&\\
		& $1p$& B& -13.31(-12.03) & -19.79(-18.31)& -24.78(-23.13)&\\
		&&$\Gamma/2$ & 9.11(9.16) & 8.58(8.67)& 8.19(8.31)&\\
		\hline
		${\rm{^{208}Pb}}$ & $1s$& B & -18.06(-16.65) &-25.23(-23.61)&-30.67(-28.88)&\\
		&&$\Gamma/2$ & 9.44(9.48) & 8.89(8.97)& 8.49(8.60)&\\
		& $1p$& B& -14.33(-12.99) & -21.16(-19.61)& -26.39(-24.67)&\\
		&&$\Gamma/2$ & 9.40(9.44) & 8.85(8.94)& 8.45(8.57)&\\
		\hline
	\end{tabular}
	\caption{ \raggedright{Contributions of neutral (charged) kaon loops to the binding energies (in MeV) $\rm{B}$ and half widths (in MeV) $\Gamma/2$ of $\phi$-mesic–nucleus bound states in various nuclei, ${\rm{^{4}He}}$,  ${\rm{^{12}C}}$, ${\rm{^{16}O}}$, ${\rm{^{40}Ca}}$, ${\rm{^{90}Zr}}$, ${\rm{^{197}Au}}$, and ${\rm{^{208}Pb}}$ for different cutoff values of $\Lambda_K$. 
			The symbol `$\times$' refers to unbound states.}}
	\label{B}
\end{table}
	However, a few of them can show up as distinct bound states.
	For instance, for $\rm{^{4}He}$, since all the obtained binding energies ( $\sim$ $-$1 or $-$2 MeV) are lesser compared to the half-width ($\sim$ 11 MeV), there will be a significant overlap of these states with the energies of the continuous energy spectrum associated with the quasi-free $\phi$ production.
	The chances to observe distinct peaks of the bound states increase towards higher values of the cutoff parameter.
	As can be read from Table \ref{B}, the $\phi$-${\rm{^{12}C}}$ states are probable for $\Lambda_K$ = 3 GeV and $\Lambda_K$ = 4 GeV.
	For heavier nuclei, the existence of several closely spaced bound states may cause the signals to be smeared out due to the overlapping of many contributions.
	Thus, another important factor of observing a clear bound state is the separation of levels $\geq \Gamma/2$.
	The states satisfying these two criteria, i.e., the relatively small half-widths, smaller than the binding energies, and smaller than the separation among levels, lead to the possible clear observation of those states.
	Thus the best chances for observing clear bound states are around the region of ${\rm{^{16}O}}$, where, only one s-bound state exists with a half width is about a factor of two smaller than the binding energy, at $\Lambda_K$ = 4 GeV.
	As mentioned in Ref. \cite{lauraplb10,oset02}, even if no distinct peaks are observed, some spectral strength may still be detected in the bound region, stretching the energy down to the $\rm{B+\Gamma/2}$ of the bound states, as depicted in Fig.\ref{be}.
	Notably, the current estimation does not consider the spin-orbit interaction.
	Thus we prefer to present the results for the ground states only.
	In comparison to the results presented in Ref. \cite{cobos17}, because of the inclusion of scalar-isovector potential, a few states are now in probable existence.
	For example, in the present scenario, at $\Lambda_K$ = 3 GeV, $\phi$-${\rm{^{12}C}}$ state is not only showing as a bound state but providing a promising candidate passing all the aforementioned criteria.
	The inclusion of such potentials modifies the binding energies as well as the absorption widths essentially.
	However, a lower ratio of binding energy to width reduces the likelihood of observing these bound states in the region of lighter nuclei.
	
\begin{figure}[th]
		\centerline{\includegraphics[width=8.2cm,height=7.3cm]{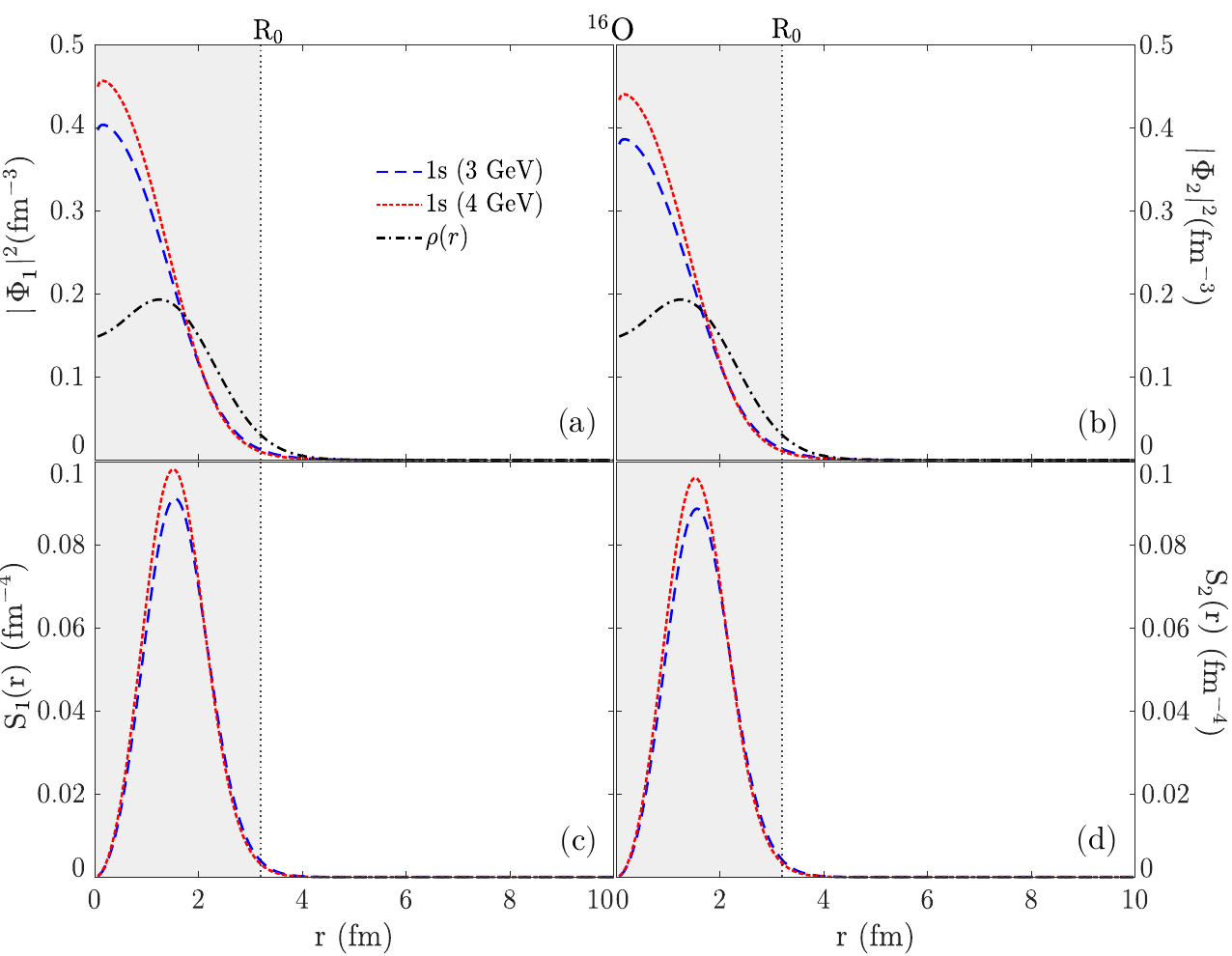}}
		\caption{\raggedright{Probability densities (a-b), nuclear density distributions(a-b) and the overlapping probabilities (c-d) for $\phi$-mesic-nuclei states in ${\rm{^{16}O}}$ nucleus, considering neutral (left panel) and charged (right panel) kaon loops. The vertical dotted lines represent the radius of the nucleus, $\rm{R_0}$ (in fm) and the shaded region implies the interior of the nucleus.}}
		\label{overO}
	\end{figure}

\begin{figure}[th]
		\centerline{\includegraphics[width=8.2cm,height=7.3cm]{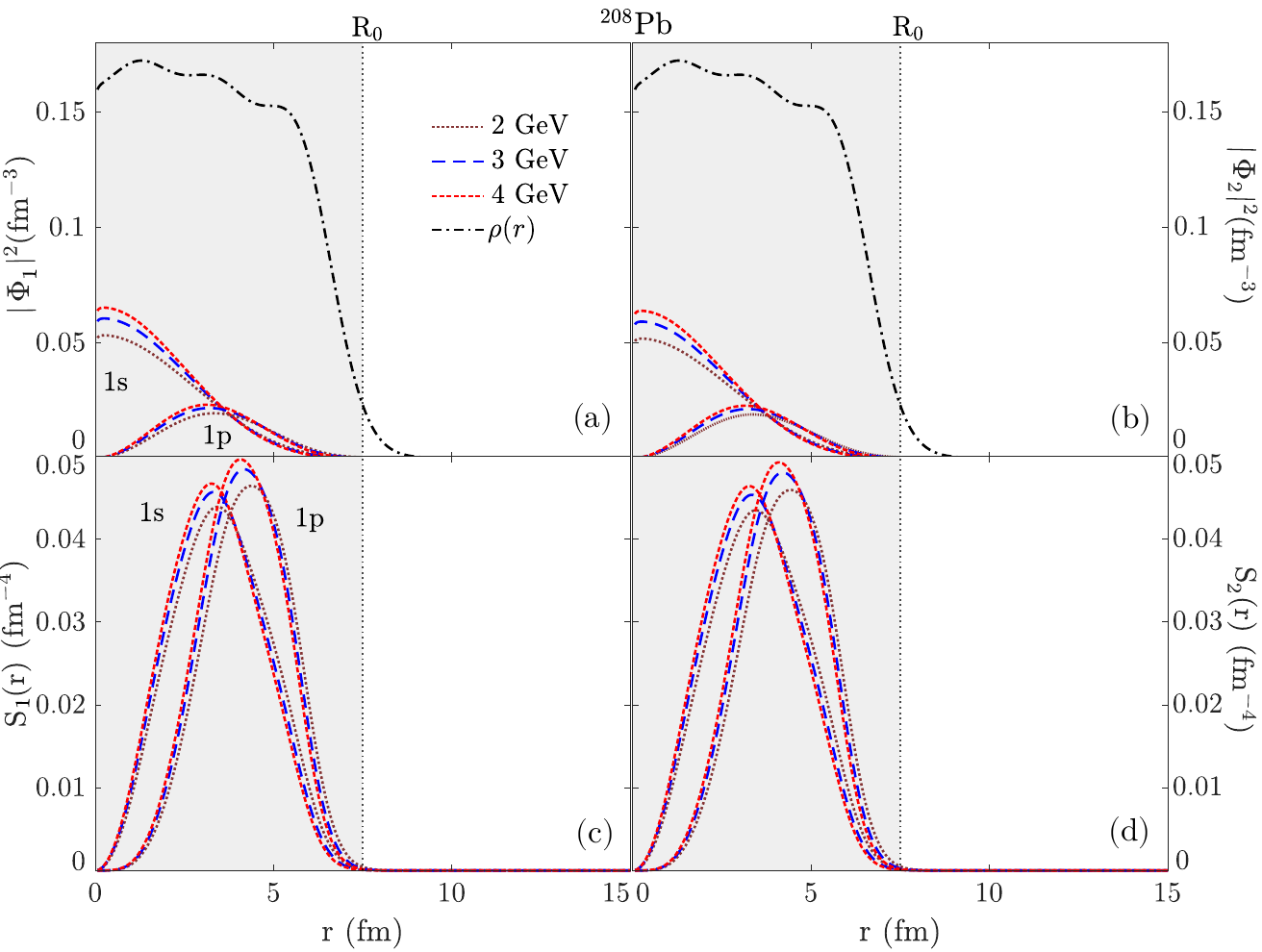}}
		\caption{\raggedright{Same as Fig.\ref{overO} for ${\rm{^{208}Pb}}$}}.
		\label{overPb}
	\end{figure}

	The feasibility of the experimental observation can be ensured by the cross-section estimation of the particular state.
	The cross-section for the reaction is proportional to $|\Phi(q)|^2$, where $\Phi(q)$ is the $\phi$-mesic-nuclei bound wave function in momentum space with $q$ as the momentum transferred to the bound $\phi$ meson.
	This suggests a low energy hadronic reaction would produce $\phi$ meson with smaller momentum.
	However, it is difficult to produce $\phi$ meson with nearly zero momentum relative to the nucleus.
	The radial dependence of the obtained $\phi$-mesic-nuclei bound state wave function, $\Phi_i(r)$, are presented in Figs.\ref{overO}-\ref{overPb} for $\rm{^{16}O}$ and $\rm{^{208}Pb}$, respectively.
	As the nucleon numbers of the atomic nuclei increase, the radii of the  nuclei become larger, and accordingly, the wave functions of the mesons become extended.
	
Next, we investigate the region where the aforementioned states are forming and the possibilities of the nuclear densities $\rho(r)$ to be probed by the bound $\phi$ meson.
Two representative nuclei, ${\rm{^{16}O}}$ and ${\rm{^{208}Pb}}$ are considered to demonstrate the patterns.
The nuclear density distributions for these nuclei are computed using the QMC model, including the coulomb interaction \cite{me24}.
Figure \ref{overO}(a,b) illustrates the probability densities of $\phi$ bound states, $(|\Phi_i(r)|^2)$, together with $\rho(r)$ and  Fig.\ref{overO}(c,d) depicts the overlapping densities, $S_i(r) = \rho(r)|\Phi_i(r)|^2 r^2$ \cite{overlapplb03}, for the region of ${\rm{^{16}O}}$ nucleus.
	Results for ${\rm{^{208}Pb}}$ are shown in Fig.\ref{overPb}.
As we see from the figures, the states are embedded inside of the nuclei, indicating their formations as mesic-nuclei bound states.
Furthermore, the ${\rm{^{16}O}}$ nucleus accommodates the shallow bound states, whereas the deeply bound mesic-nuclei states are obtained in ${\rm{^{208}Pb}}$ nucleus.
Notably, as we discussed, we have disregarded the states where the absolute values of the binding energies are much smaller than the corresponding half-widths.
The obtained overlapping densities peak nearly around the central densities, which is largely independent of the nucleus and the quantum numbers, indicating the mesic-nuclei bound states can predominately probe the central nuclear density.
	 
\section{SUMMARY}
\label{summary}
We have investigated the properties of the $\phi$ meson within strongly interacting symmetric as well as asymmetric nuclear matter and atomic nuclei, using the tree level $\phi K^0\bar{K}^0$ and $\phi K^+ K^-$ Lagrangian densities with the in-medium masses of $K$ and $\bar{K}$ mesons as calculated within the QMC model.
We studied the influence of medium effects on the vector meson ($\phi$) production cross-sections due to the scattering of $K^0\bar{K}^0$  as well as  $K^+ K^-$ within both symmetric and asymmetric nuclear matter by implementing the in-medium Breit-Wigner spectral function from the $\phi$ meson self-energies. 
The real and imaginary parts of the self energies are related to the mass and the partial decay widths of the $\phi$ meson, which are calculated for a range (2-4 GeV) of cutoff parameter used in the regularization of the self-energy and, by employing the medium-modified masses of the $K(K^+, K^0)$ and $\bar{K}(\bar{K}^0, K^-)$ mesons of our previous study in Ref. \cite{me23}.
In the QMC model, a direct interaction between quarks and mean meson fields modifies the open flavor meson's properties self consistently.
The findings of the current study can be summarized as follows:
\
At finite baryon densities, we observe a decrease (increase) in the $\phi$ mass (partial as well as total decay widths) within the nuclear matter, irrespective of the cutoff parameter $\Lambda_K$, which shows, in general, a decreasing peak position as well as a drop in peak height towards higher densities.
Compared to SNM, in ANM, the $\phi$ mass receives a comparatively lesser mass drop and decay width, leading to significant modifications of the peaks in the higher density region.
The difference between the peaks obtained for both channels is moderate with a comparatively lower peak position of the neutral channel than the charged channel.
As $\Lambda_K$ increases, the production cross-section peaks for each density become more distinct with the pronounced isospin asymmetry effect and with a greater peak height.
Such shape change in $\phi$ peaks indicates the medium modifications of its properties, which can be probed with a high-resolution beam.

The behavior of the ratio of the branching ratios of the decay channels within the nuclear matter shows that the ratio decreases with increasing density, and towards higher density it saturates to unity, indicating comparable branching ratios of both the neutral and charged $K\bar{K}$ channels, irrespective of the cutoff values.

As a consequence of the increasing total $\phi$ decay width, the lifetime of the $\phi$ becomes substantially low.
Furthermore, we have observed a downward shift of the spectral peak towards lower invariant masses.
These indicate the $\phi$ meson to decay within the nuclear dimension and the probable formation of bound states of $\phi$ with atomic nuclei.
The $\phi$ mesic nuclei are considered to be very important objects to study the medium modification of the $\phi$-meson spectral properties at finite density.
In the present work, we have studied the probable formation of the $\phi$ mesic nuclei and discussed the feasibility of observing such states.
We have calculated the binding energies and the absorption widths of the $\phi$ meson bound states for different values of $\Lambda_K$, considering two prominent kaon channels, by solving the Klein-Gordon equation with the $\phi$-potentials within the ${\rm{^{4}He}}$, ${\rm{^{12}C}}$, ${\rm{^{16}O}}$, ${\rm{^{40}Ca}}$, ${\rm{^{90}Zr}}$, ${\rm{^{197}Au}}$, and ${\rm{^{208}Pb}}$ nuclei.
We studied the sensitivity of these expected bound states on the mass number and the cutoff parameter.  
A precise measurement of these binding energies would significantly insights into the meson-nucleus interactions, the dynamics of strongly interacting
systems, and the nature of these states.
The stronger the bound state's strength, the closer the meson stays at the nuclei, letting these states probe even smaller variations in the nuclear medium effects.
Depending upon the observed results, the $\phi$-bound states in light nuclei might have better opportunities of being detected.
However, a further investigation of the reaction cross-section can only offer the feasibility for the detection of these states.
These findings are relevant for the experimental investigation of the strong interaction in the low-energy regime in understanding the low-energy meson-nucleon interaction with implications in diverse fields, from the search for exotic mesic nuclear bound states to the structure of compact astrophysical objects like neutron stars.
The upcoming ${\rm{\bar{P}}ANDA}$ at FAIR, J-PARC, and JLab experiments will be particularly significant for such studies.

\begin{thebibliography}{99} 
	\bibitem{pdg}  
	R. L. Workman \textit{et al.} (Particle Data Group), Prog. Theor. Exp. Phys. {\bf 2022}, 083C01 (2022).
		\bibitem{kek}
		R. Muto \textit{et al.} [KEK-PS-E325 collaboration], Phys. Rev.	Lett. {\bf 98}, 042501 (2007).
		\bibitem{alice}
		Acharya, S., et al., Phys. Rev. Lett. {\bf 127}, 172301 (2021).
		\bibitem{e16exp}
		K. Aoki, \textit{et al.}, Few Body Syst. \textbf{64}, 63 (2023).
		\bibitem{leps}
		T. Ishikawa \textit{et al.}, Phys. Lett. B {\bf 608}, 215 (2005).
		\bibitem{clas}
		M. H. Wood \textit{et al.} [CLAS collaboration], Phys. Rev. Lett. {\bf 105}, 112301 (2010).
		\bibitem{anke}
		A. Polyanskiy, et al., Phys. Lett. B {\bf 695}, 74 (2011).
		\bibitem{hades_model}
		E.Y. Paryev, Nucl. Phys. A {\bf 1032}, 122624 (2023).
		\bibitem{hades}
		J. Adamczewski-Musch et al. (HADES collaboration), Phys. Rev. Lett. {\bf 123}, 022002 (2019).
		\bibitem{e16}
		S. Yokkaichi, \textit{et al.}, P16, http://j-parc.jp/researcher/Hadron/en/pac0606/pdf/p16-Yokkaichi 2.pdf (2006).
		\bibitem{e88}
		H. Sako, \textit{et al.}, P88, http://j-parc.jp/researcher/Hadron/en/pac2107/pdf/P88 2021-12.pdf (2022).
		\bibitem{e29}
		H. Ohnishi, \textit{et al.}, P29, http://j-parc.jp/researcher/Hadron/en/pac1007/pdf/KEK J-PARC-PAC2010-02.pdf (2010).
		\bibitem{me23}
		A. Mondal and A. Mishra, Phys. Rev. C {\bf 109}, 025201 (2024).
		\bibitem{alice_hal}
		E. Chizzali, Y. Kamiya, R. Del Grande, T. Doi, L. Fabbietti, T. Hatsuda, Y. Lyu, Phys. Lett. B {\bf 848}, 138358 (2024).
		\bibitem{paryev}
		E.Ya. Paryev, Nucl. Phys. A {\bf 1032}, 122624 (2023).
		\bibitem{e26}
		K. Aoki, for the J-PARC E16 Collaboration, Contribution to the Xth Workshop on Particle Correlations and Femtoscopy (WPCF2014), arXiv:1502.00703 [nucl-th].
		\bibitem{jlab}
		M. Paolone, Z. E. Meziani, \textit{et al.}, https://www.jlab.org/exp prog/PACpage/PAC42/PAC42 FINAL Report.pdf (2014).
		\bibitem{kreinplb11}
		G. Krein, A. W. Thomas, and K. Tsushima, Phys. Lett. B {\bf 697}, 136 (2011).
		\bibitem{Ashok}
		A.~Das, \textit{Lectures on quantum field theory}, World Scientific, 10.1142/6938 (2008).
		\bibitem{Mishra21}
		A. Mishra and S. P. Misra, Eur. Phys. J. A \textbf{57}, 98 (2021).
		\bibitem{Mishra24}
		A. Mishra, A. Kumar, and S. P. Misra, Phys. Rev. D \textbf{110} 014003 (2024).
		\bibitem{me24}
		A. Mondal and A. Mishra, Phys. Rev. C {\bf 110}, 055201 (2024).
		\bibitem{cobos17}
		J. J. Cobos-Mart\'{i}nez, K. Tsushima, G. Krein, and A. W. Thomas, Phys. Rev. C {\bf 96}, 035201 (2017).
		\bibitem{cobos17plb}
		J. J. Cobos-Mart\'{i}nez, K. Tsushima, G. Krein, and A. W. Thomas, Phys. Lett. B {\bf 771}, 113 (2017).
		\bibitem{leinweberprd01}
		D. B. Leinweber, A. W. Thomas, K. Tsushima, and S. V. Wright, Phys. Rev. D {\bf 64}, 094502 (2001).
		\bibitem{klingl97}
		F. Klingl, N. Kaiser, and W. Weise, Nucl. Phys. A {\bf 624}, 527 (1997).
		\bibitem{klingl98}
		F. Klingl, T. Waas, and W. Weise, Phys. Lett. B {\bf 431}, 254 (1998).
		\bibitem{yasuiprd09}
		S. Yasui and K. Sudoh, Phys. Rev. D \textbf{80}, 034008 (2009).
		\bibitem{zemiepja21}
		G. N. Zeminiani, J. J. Cobos-Mart\'{i}nez, and K. Tsushima, Eur. Phys. J. A \textbf{57}, 259 (2021).
		\bibitem{rajesh20}
		R. Kumar, A. Kumar, Phys. Rev. C {\bf 102}, 045206 (2020).
		\bibitem{ko92}
		C. M. Ko, P. Levai, X. J. Qiu, and C. T. Li, Phys. Rev. C {\bf 45}, 1400 (1992).
		\bibitem{oset01}
		E. Oset, A. Ramos, Nucl. Phys. A {\bf 679}, 616 (2001).
		\bibitem{santos}
		A. M. Santos, P. K. Panda, and C. Provid$\rm{\hat{e}}$ncia, Phys. Rev. C {\textbf{79}}, 045805 (2009).
		\bibitem{liu}
		B. Liu, V. Greco, V. Baran, M. Colonna, and M. Di Toro, Phys. Rev. C {\bf 65}, 045201 (2002).
		\bibitem{tsushimaplb98}
		K. Tsushima, K. Saito, A.W. Thomas, and S.V. Wright, Phys. Lett. B {\bf 429}, 239 (1998).		
		\bibitem{oset02}
		C. Garcia-Recio, J. Nieves, T. Inoue and E. Oset, Phys. Lett. B \textbf{550}, 47 (2002).
		\bibitem{lauraplb10}
		C. Garc\'{i}a-Recio, J. Nieves b, L. Tolos, Phys. Lett. B {\bf 690}, 369 (2010).	
		\bibitem{overlapplb03}
		Toshimitsu Yamazaki, and Satoru Hirenzaki, Phys. Lett. B {\bf 557}, 20 (2003).			
\end{thebibliography}
%

    \end{document}